# Combined Effects Due to Phase, Intensity, and Contrast in Electrooptic Modulation: Application to Ferroelectric Materials

L. Guilbert, J. P. Salvestrini, H. Hassan, and M. D. Fontana

*Abstract*— The combination of phase, intensity, and contrast effects during electrooptic modulation is theoretically and experimentally investigated. One consequence of this combination is the modification of the amplitude of the single-frequency signals which are commonly used as working points for electrooptic modulators and for the measurements of the electrooptic coefficients. Another consequence of direct intensity modulation is to shift the double-frequency points of the transfer function from the positions they normally occupy at the intensity extrema. They can even make them disappear if the direct intensity modulation is stronger than the phase modulation. Such phenomena are expected with any ferroelectric material in which a significant part of the incident light is deflected or scattered by domain walls or grain boundaries. They can lead to considerable mistakes in the determination of the electrooptic coefficients. Appropriate procedures to extract the different contributions are explained. Experimental results in rubidium hydrogen selenate are given, and consequences of the working of electrooptic modulators are discussed.

## I. INTRODUCTION

IN MANY electrooptic (EO) applications, the modulation of light is usually based on the Pockels effect (first-order EO effect) or on the Kerr effect (second-order EO effect). The best EO materials presenting these effects are most often single crystals, either paraelectric or ferroelectric, but preferably free from domains or lattice defects, especially for EO devices working in laser beam treatment and requiring a perfect optical quality. For less demanding applications, other materials can be used, such as liquid crystals or ferroelectric ceramics. In the latter (PZT or PLZT) and, as a rule, in any material having domain structures with both ferroelectric and ferroelastic properties, specific effects related to domain dynamics may appear: not only the phase shift, but also the intensity or the contrast of the transmitted light can be modified or modulated by an external electric field.

Recently, we evidenced a giant EO effect related to domain dynamics in rubidium hydrogen selenate (RHSe) at low frequency [1]–[3] (0–100 kHz). The EO coefficients involved in this effect have been measured with a classical Sénarmont's setup. During the experiments under an ac field, it was observed that a direct intensity modulation and/or a contrast modulation were sometimes superimposed onto the phase modulation. Similar effects have been also evidenced in semitransparent PLZT samples. They manifest themselves by a shift of the double-frequency points and by a modified amplitude of the single-frequency signal measured at the middle point of the transfer function. These phenomena—if ignored—can lead to erroneous results when determining the EO coefficients. Moreover, in the case of EO modulators devoted to applications in intensity modulation, it could be important to choose conveniently the linear working point of the modulator in order to enhance the overall amplitude of the signal.

This paper deals with combined modulation effects—phase, intensity, and contrast—in the Sénarmont setup. The convenient procedures to extract the different contributions are explained. They can be relevant for EO measurements performed on multidomain ferroelectric materials, or on other kinds of materials exhibiting both the Pockels effect and electroabsorption. The frequency dispersions that we obtained this way for the phase, intensity, and contrast modulation coefficients in the RHSe crystal are given as illustrative results. Also, the consequences of the combined modulation effect for EO modulators are analyzed.

## II. SENARMONT'S SETUP

The classical Sénarmont's setup commonly used for EO measurements is shown in Fig. 1. The sample is placed between a polarizer and a quarter-wave plate, the neutral axes of which are oriented at 45° from the axes of the crystal and the polarizer. This setup allows one to obtain at the output of the quarter-wave plate a linear polarization, the direction of which depends upon the phase shift introduced by the crystal between the two components of the lightwave polarization. After the quarter-wave plate, a rotating analyzer allows one to measure the variations of the phase shift induced by the applied electric field or by other external factors, such as a mechanical stress or a temperature variation.

If the crystal or any other element of the setup is not optically perfect, the contrast is generally not equal to unity. The transfer function of the light intensity transmitted through the setup (Fig. 2) can be written as

$$i(\phi) = \frac{I_M}{2}(1 - \gamma \cos\phi) \quad (1)$$

where $\gamma$ ($< 1$) is the contrast, $I_M$ the maximal intensity which could be transmitted if the contrast were equal to unity,

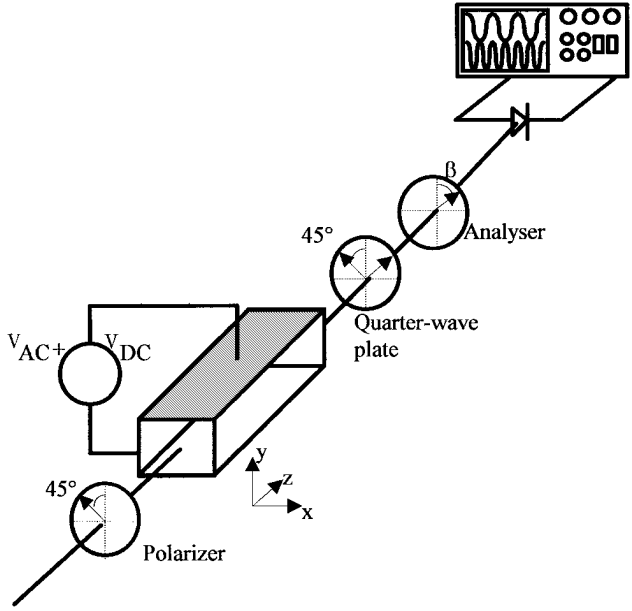

Fig. 1. Sénarmont's setup commonly used for electrooptic measurements. The axes of the polarizer and the quarter-wave plate are set at 45° from the neutral lines of the crystal in order to obtain a quasi-linear polarization of the lightwave at the input of the analyzer.

and $\phi/2$ is the angle between the analyzer and the linear polarization of the output lightwave

$$\phi/2 = \beta - \Gamma/2. \quad (2)$$

$\beta$ is the angular position of the analyzer and $\Gamma$ is the phase shift introduced by the EO crystal. Usually, the maximal intensity $I_M$ and the contrast $\gamma$ are considered constant parameters, and only the phase shift $\Gamma$ is supposed to be sensitive to the applied electric field $E$

$$\Gamma(E) = (2\pi L/\lambda)\Delta n(E) \quad (3)$$

where $L$ is the length of the crystal, $\lambda$ is the wavelength, and $\Delta n(E)$ is the field-sensitive birefringence of the EO sample

$$\Delta n(E) = \Delta n_0 - \frac{1}{2}n^3 r_{\text{eff}} E \quad (4)$$

where $\Delta n_0$ is the natural birefringence and $n^3 r_{\text{eff}}$ is the effective EO coefficient to be measured.

Since the derivative of (1) versus $E$ is equal to zero at the point $M_0$ ($M_0'$) where the transmitted intensity is minimal (maximal), these points correspond to the so-called *double-frequency points*: an ac field of frequency $\omega$ applied to the crystal yields an optical signal modulated at frequency $2\omega$. The double-frequency points are commonly used to determine the static EO coefficient [4]; as soon as a step of dc field $\Delta E$ is superimposed on the ac field, the double-frequency signal is lost and the analyzer has to be rotated by an angle $\Delta\beta = \Delta\Gamma/2 = (\pi L/2\lambda)n^3 r_{\text{eff}}\Delta E$ to regain the double-frequency signal.

On another hand, the so-called *middle point* $M_1$ (or $M_1'$) corresponding to the medium intensity $I_M/2$ of the transfer function (Fig. 2) can be used to determine the EO coefficient at any frequency. Measuring the peak-to-peak amplitude $i_{\text{pp}}$ of the modulated signal at one of these working points ($\phi =$

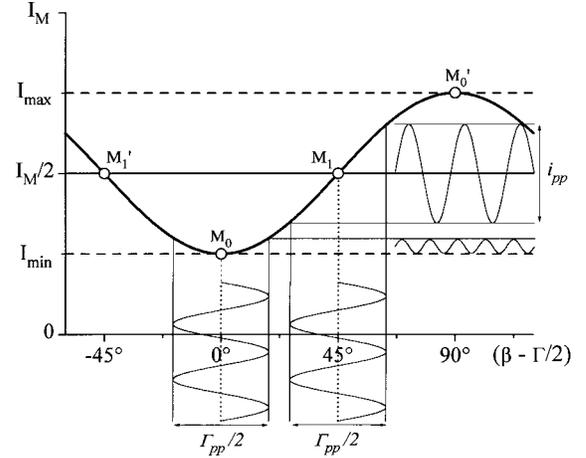

Fig. 2. Transfer function $i(\beta)$ through Sénarmont's setup.

$\pm\pi/2$), one obtains immediately from the above equations

$$\bar{n}^3 r_{\text{eff}}(\omega) = \frac{2\lambda}{\pi L E_{\text{pp}}(\omega)} \arcsin\left(\frac{i_{\text{pp}}(\omega)}{I_{\max} - I_{\min}}\right)$$
$$\approx \frac{2\lambda}{\pi L E_{\text{pp}}(\omega)} \frac{i_{\text{pp}}(\omega)}{I_{\max} - I_{\min}} \quad \text{for small signals} \quad (5)$$

where $I_{\max} - I_{\min} = \gamma I_M$ is the intensity range of the transfer function and $E_{\text{pp}}$ the peak-to-peak amplitude of the applied ac field at frequency $\omega$. In (5), the dimensionless ratio

$$\frac{i_{\text{pp}}}{I_{\max} - I_{\min}}$$

is commonly noted $m$ and called the "modulation factor" or the "ratio of phase modulation."

Typical experiments using single-domain crystals in the Sénarmont's setup generally allow accurate measurements of the effective EO coefficient *provided that the modulated signal is due to phase modulation only*. However, in the case of multidomain crystals or ceramics, it is sometimes observed that the maximal transmitted intensity $I_M$ is also directly modulated by the ac field. Moreover, the contrast $\gamma$ of the transfer function can be modulated also. As we shall see in Section IV, the domain structure of the sample is responsible for these phenomena. The conjugation of several modulation effects can lead to unexpected features and to possible mistakes in the determination of the EO coefficients.

### III. COMBINED MODULATION EFFECTS

*A. Theory*

Let us now assume that in (1) the phase shift, the maximal intensity, and the contrast can be modulated together at the same frequency $\omega$ by the applied ac field. The derivative of (1) versus electric field $E$ yields

$$\frac{di}{dE} = \frac{\gamma I_M \sin\phi}{2}m_\phi - \frac{\gamma I_M \cos\phi}{2}m_\gamma + \frac{I_M(1 - \gamma\cos\phi)}{2}m_I \quad (6)$$

where $m_\phi = d\phi/dE = -d\Gamma/dE$ is the phase modulation coefficient, $m_\gamma = (1/\gamma)\cdot(d\gamma/dE)$ is the contrast modulation coefficient, and $m_I = (1/I_M)\cdot(dI_M/dE)$ is the intensity modulation coefficient ($m_\gamma$ and $m_I$ are expressed in m/V, $m_\phi$ in rad·m/V). According to this definition, the above-defined ratio of phase modulation $m$ (dimensionless) is simply given by the product $m = m_\phi \cdot E_{\rm pp}$.

If only the phase modulation is present ($m_\phi \neq 0, m_\gamma = m_I = 0$), one obtains the classical results described in Section II: the double-frequency points are located at the extrema $\phi = k\pi$ of the transfer function. The determination of the effective EO coefficient is straightforward from (5).

If only the intensity modulation is present ($m_I \neq 0, m_\gamma = m_\phi = 0$), there should be no double-frequency points in the transfer function: the amplitude of the modulated signal should increase monotonically, without changing phase, as the working point is moved from the minimum $M_0$ to the maximum $M_0'$ by rotating the analyzer.

If only the contrast modulation is present ($m_\gamma \neq 0, m_\phi = m_I = 0$), one should expect a double-frequency signal at the middle points $M_1$ and $M_1'$ of the transfer function. The extremal points $M_0$ and $M_0'$ should correspond to a maximal amplitude of the single-frequency modulation, with opposite phases of the signal at $M_0$ and $M_0'$. In other words, the characteristics of the contrast modulation on the transfer function are exactly reversed to the ones of phase modulation.

When several modulation effects are combined simultaneously, a first consequence is that the double-frequency points—if they exist—are shifted from their usual positions $M_0$ and $M_0'$. *If one assumes that all these modulation effects are originated from a same physical cause (e.g., domain dynamics in the case of multidomain materials), the modulation coefficients $m_\phi, m_\gamma, m_I$ in (6) can be considered as real quantities*, either positive or negative, depending on the relative phases (either equal or opposite) of the physical effects. The new positions of the double-frequency points with respect to the minimal point $M_0$ can be easily calculated, as they correspond to the values of $\phi$ which annul (6)

$$\Delta\beta_{2\omega}^\pm = \arctan\left(\frac{-m_\phi \pm \sqrt{\Delta}}{m_\gamma + m_I(1 + 1/\gamma)}\right). \quad (7)$$

The general criterion for the double-frequency points to exist is

$$\Delta = m_\phi^2 + m_\gamma^2 + 2m_\gamma m_I - m_I^2(1/\gamma^2 - 1) > 0. \quad (8)$$

- If intensity modulation is absent ($m_I = 0$), this criterion is always satisfied, and the position of the double frequency point is simply given by:

$$(\Delta\beta_{2\omega})_{m_I=0} = \frac{1}{2}\arctan\left(\frac{m_\gamma}{m_\phi}\right). \quad (9)$$

- If contrast modulation is absent ($m_\gamma = 0$), the criterion (8) is simplified to

$$\left|\frac{m_\phi}{m_I}\right| > \frac{\gamma}{\sqrt{1-\gamma^2}} \quad (10)$$

and, if it is satisfied, the double-frequency points are located at

$$(\Delta\beta_{2\omega}^\pm)_{m_\gamma=0} = \arctan\left(\frac{-\gamma m_\phi \pm \sqrt{\gamma^2 m_\phi^2 - (1-\gamma^2)m_I^2}}{(\gamma+1)m_I}\right). \quad (11)$$

A second consequence of the combined modulation effects is that the amplitudes of the modulated signal at the middle points $M_1$ and $M_1'$ of the transfer function are not equal. For small signals ($|i_{\rm pp}| \ll \gamma I_M = I_{\max} - I_{\min}$), these amplitudes $i_{\rm pp}^+$ and $i_{\rm pp}^-$ are given by

$$\frac{i_{\rm pp}^\pm}{I_{\max} - I_{\min}} = \frac{1}{2}\left(\frac{m_I}{\gamma} \pm m_\phi\right)E_{\rm pp}. \quad (12)$$

Equation (12) shows that the discrepancy between the two signals is due to intensity modulation only. The contrast modulation does not affect their amplitude. At the point $M_1$ ($\phi = +\pi/2$), the combination of phase and intensity modulations is additive and the signal is enhanced (if $m_\phi$ and $m_I$ have the same sign). At the point $M_1'$ ($\phi = -\pi/2$), the combination is subtractive and the signal is weakened.

### B. Practice: How to Extract the Different Contributions

Experimentally, the first thing to do is to measure the contrast $\gamma$ of the transfer function (preferably when the ac field is not applied). One obtains the extremal intensities at the points $M_0$ and $M_0'$, the output analyzer being rotated by 90° in between. The contrast is then estimated by

$$\gamma = \frac{I_{\max} - I_{\min}}{I_{\max} + I_{\min}} = \frac{V_{\max} - V_{\min}}{V_{\max} + V_{\min} - 2V_0} \quad (13)$$

where $V_0$ is the dark signal and $V_{\max}$ and $V_{\min}$ are the extremal dc signals given by the photodetector.

As soon as an ac field is applied to the sample, it is easy to check whether this field is responsible for a direct intensity modulation: one has just to remove the output analyzer from the setup. If any modulated signal $i_{\rm pp}^0(\omega)$ is observed without an analyzer, the intensity modulation coefficient $m_I(\omega)$ can be directly measured

$$m_I(\omega) = \frac{1}{I_M^0}\frac{i_{\rm pp}^0(\omega)}{E_{\rm pp}(\omega)} \quad (14)$$

where $I_M^0$ is the average (dc) value of the transmitted intensity (without analyzer). It is useful to observe whether the optical signal is in phase or in opposition with respect to the electrical signal: this determines the sign (+ or − respectively) of the modulation coefficient $m_I$.

The analyzer can now be replaced in the setup and rotated by 45° on both sides of the position $\beta_0$ corresponding to the minimal intensity. The amplitudes $i_{\rm pp}^+$ and $i_{\rm pp}^-$ of the optical signal at the points $M_1$ and $M_1'$ are then measured, *as algebraic quantities*, bearing the convenient sign + or − depending on whether the optical signal is in phase or in opposition with the electrical signal. So, one can deduce the phase modulation coefficient $m_\phi$ from the *difference* $i_{\rm pp}^+ - i_{\rm pp}^-$, according to (12)

$$m_\phi(\omega) = \frac{1}{I_{\max} - I_{\min}}\frac{i_{\rm pp}^+(\omega) - i_{\rm pp}^-(\omega)}{E_{\rm pp}(\omega)}. \quad (15)$$

One can also deduce the intensity modulation coefficient $m_I$ from the algebraic sum $i_{\rm pp}^+ + i_{\rm pp}^-$, provided that the contrast $\gamma$ has been measured

$$m_I(\omega) = \frac{\gamma}{I_{\max} - I_{\min}}\frac{i_{\rm pp}^+(\omega) + i_{\rm pp}^-(\omega)}{E_{\rm pp}(\omega)}. \quad (16)$$

If one observes that the phase of the optical signal is the same at both points $M_1$ and $M_1'$, it means that intensity modulation prevails over phase modulation. In other words, one has $|m_I| > \gamma |m_\phi|$ in (12). This can happen especially when the contrast is poor. In this case, the determination of the phase modulation coefficient given by (15) is generally not very accurate, but there is no other way to do it.

Inversely, if one observes that the optical signal has opposite phases at the middle points $M_1$ and $M_1'$, phase modulation prevails over intensity modulation. The phase modulation coefficient $m_\phi$ can be deduced accurately from (15), but the intensity modulation coefficient $m_I$ should better be obtained by the direct measurement without analyzer, from (14), rather than from (16).

The effective EO coefficient $n^3 r_{\mathrm{eff}}(\omega)$ can always be deduced from (5), provided that the peak-to-peak amplitude $i_{\mathrm{pp}}$ of the optical signal is replaced by the half-difference $(i_{\mathrm{pp}}^+ - i_{\mathrm{pp}}^-)/2$ of the algebraic amplitudes measured at the middle points $M_1$ and $M_1'$.

It should be stressed that when any direct intensity modulation is superimposed to the phase modulation, a classical measurement of the modulated signal at only one of the middle points can lead to a significant error in the determination of the EO coefficient, either by excess or by default, depending on the middle point ($M_1$ or $M_1'$) chosen for the measurement. The relative error $m_I/\gamma m_\phi$ can be large when the contrast is poor. The double measurement at both points $M_1$ and $M_1'$ allows one to eliminate this error.

If one is interested in the determination of the contrast modulation coefficient $m_\gamma$, this can be done from (7), provided that the double-frequency points can be clearly observed and accurately located relatively to the point $M_0$ corresponding to the minimal intensity. If $m_\phi$, $m_I$, and the positions $\Delta\beta_{2\omega}^\pm$ of the double-frequency points have been measured, one obtains from (7)

$$m_\gamma = -\frac{2 m_\phi}{\tan\Delta\beta_{2\omega}^+ + \tan\Delta\beta_{2\omega}^-} - m_I(1 + 1/\gamma). \qquad (17)$$

Note that when intensity modulation is absent ($m_I = 0$), it should be observed that the double-frequency points $M_{2\omega}$ and $M_{2\omega}'$ are shifted identically, with respect to the minimal point $M_0$ and to the maximal point $M_0'$, respectively. Accordingly, any difference between the shifts of the two double-frequency points is another signature of a direct intensity modulation.

## IV. APPLICATION TO FERROELASTIC CRYSTALS

Combined modulation effects can be observed at least in two different kinds of materials:

- Materials displaying both the Pockels effect (field-induced birefringence) and the Franz–Keldysh effect (field-induced absorbance): in particular, semiconducting crystals studied at wavelengths close to the gap-related absorption band. These materials will not be considered hereafter, but the above calculations can apply to them, provided that the phase retardations associated with each one of the modulation effects are small or similar.
- Ferroelectric materials—crystals or ceramics—in which the domain structures have ferroelastic properties—a *sine qua non* condition for domain-related EO properties. We shall deal with this kind of materials in what follows. In particular, the direct intensity modulation induced by the deflection of light will be considered. The properties of phase modulation related to domain dynamics are explained in other publications [2], [3].

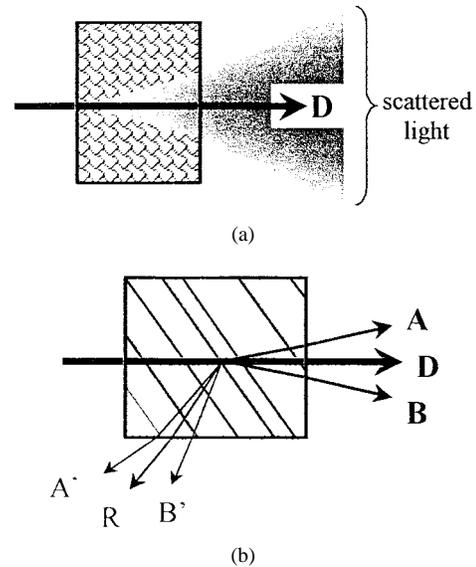

Fig. 3. Transmission of light through two kinds of microstructure in electrooptic materials. (a) Scattering by domain walls or grain boundaries in a ceramic. (b) Deflection by a layered domain structure in a single crystal. $A$, $B$: refractive transmissions; $D$: nonrefractive transmission (direct beam); $A'$, $B'$: refractive reflections; $R$: nonrefractive reflection. *In both materials, only the $D$ beam is coherent and sensitive to phase modulation.*

### A. Deflection of Light by Ferroelastic Domain Structures

Through ferroelastic domain structures, the transmitted light generally consists of a direct beam and of several deflected beams resulting from refraction and reflection processes at domain walls [5]–[7]. The angles and the intensities of the deflected beams depend on the birefringence, on the orientation of the optical indicatrix, and on the direction of propagation with respect to the domain walls. In ceramics, since the domain walls are randomly oriented, the deflection phenomenon leads to a widely scattered incoherent light, distributed in a cone around the direct beam [Fig. 3(a)]. Under an applied electric field, only the direct beam is sensitive to phase modulation, but at the same time the ac field generally induces some changes in the microstructure[1] and thus modulates the intensity ratio between the direct beam and the scattered light. Consequently, if the scattered light is hidden by a circular diaphragm at the output of the sample, one can expect a combination of intensity modulation and phase modulation on the direct beam. On the other hand, if both the direct beam and the scattered light are collected by a lens onto the photodetector, no intensity modulation will appear but the contrast will be weaker and generally modulated by the electric field.

Similar phenomena can be observed with ferroelectric crystals exhibiting a layered domain structure. In such crystals, where the domain walls are all of the same kind and parallel to each other, the deflection phenomenon gives—as a rule—six beams [7], [8], schematically shown in Fig. 3(b). Only the direct beam $D$ can be fully—or, more often, partially—coherent. The deflected beams $A$ and $B$, which are produced by the refractive transmission processes at domain walls (from low-to-high index and from high-to-low, respectively), are generally incoherent[2] because the parallel domain walls are usually spaced at random inside the crystal. The beams $A'$, $B'$ (refractive reflections), and $R$ (nonrefractive reflection) are symmetrical to $A$, $B$, and $D$ with respect to the domain walls. Most often, these reflected beams are weak as soon as the direct beam makes a sufficient angle with the plane of the domain walls. This was the case in our experiments. Their intensities will be neglected in comparison with the total transmitted intensity $I_D + I_A + I_B = I$.

---
[1]Grain boundaries can also be involved in the light scattering, but the corresponding intensity modulation should be at double frequency, since this effect is mainly related to electrostriction.

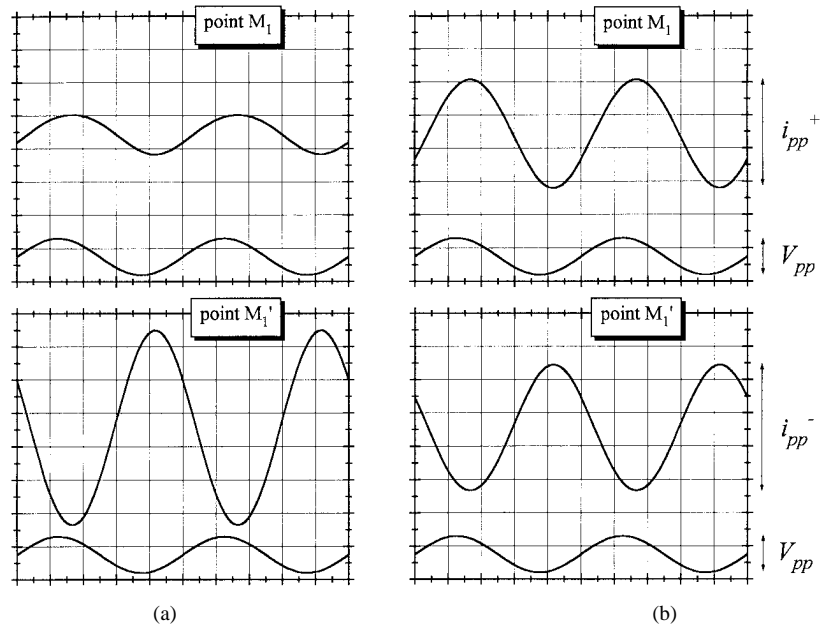

Fig. 4. Single-frequency signals recorded at the middle points $M_1$ and $M'_1$. (a) First experiment: the large difference between the peak–peak amplitudes of the optical signals is due to the direct intensity modulation on the $D$-beam. (b) Second experiment: the difference is much smaller because intensity modulation is weaker. (Recorded in RbHSeO$_4$ at 25 Hz, $E_{pp} = 100$ V/cm for both experiments.)

### B. Consequences of the Deflection in Sénarmont's Experiments

When the ac field is large enough to induce domain reversals, the sharing of the transmitted intensity $I$ between the direct beam and the deflected beams is modulated. During the negative half-periods of the field (i.e., $\mathbf{E}_{ac}$ antiparallel to the remanent polarization $\mathbf{P}_R$), new domain walls are created in the crystal and the deflected intensity $I_A + I_B$ increases to the detriment of $I_D$. During the positive half-periods of the field ($\mathbf{E}_{ac}$ directly parallel to $\mathbf{P}_R$), some domain walls are removed and $I_A + I_B$ weakens, to the benefit of $I_D$.

This effect is particularly pronounced in rubidium hydrogen selenate, up to frequencies of several tens of kilohertz, because the ferroelectric domain structure of this crystal is soft enough to be easily reversible by relatively weak ac fields (>50 V/cm peak-to-peak [2], [3]). In Sénarmont's experiments, the deflection of light leads to different consequences, depending on whether the total transmitted intensity $I = I_D + I_A + I_B$ is focalized onto the photodetector or only the direct beam $D$ selected by a diaphragm. We shall consider both experiments, performed on RHSe. The physical and optical characteristics of this crystal are published elsewhere [8].

[2]By "incoherent," we mean that the phases of the different rays in the deflected beams are widely distributed, so that these beams yield zero contrast through the setup.

*1) First Experiment: Direct Beam D Alone:* In this case, the contrast is good ($\gamma_D = 90\%$), but a significant intensity modulation is observed due to the deflection phenomenon. As can be seen in Fig. 4, the modulated signals recorded at the middle points $M_1$ and $M'_1$ of the transfer function do not have the same amplitude. Nevertheless, the two signals have opposite phases, indicating that phase modulation prevails over intensity modulation. Using (15) and (16), the modulation coefficients $m_\phi$ and $m_I$ can be determined from the amplitudes $i^+_{pp}$ and $i^-_{pp}$ measured in Fig. 4. The intensity modulation coefficient $m_I$ can also be measured directly without analyzer, from (14). We have reported in Fig. 5(a) the frequency dispersions of both $m_\phi$ and $m_I$, the latter being determined by both methods [see (14) and (16)]. The agreement between the two methods is good up to 100 Hz. The slight discrepancy observed above this frequency can be attributed to the fact that the contribution of the domain walls is collapsing, which is probably the main factor involved in the direct intensity modulation, is no longer predominant in the phase modulation: domain vibrations and lattice and ionic contributions become relatively more important in the Pockels effect with smaller dissipation effects. Thus, the ratio $m_\phi/m_I$ is no longer a real quantity because the phase retardations in the two effects are no longer equal, so the optical signal is distorted from the sinus shape. Consequently, the determination of $m_I$ is certainly less reliable from the indirect method [see (16)] than from the direct one [see (14)]. The determination of $m_\phi$ itself from (15) is probably not very accurate, especially above 100 Hz.

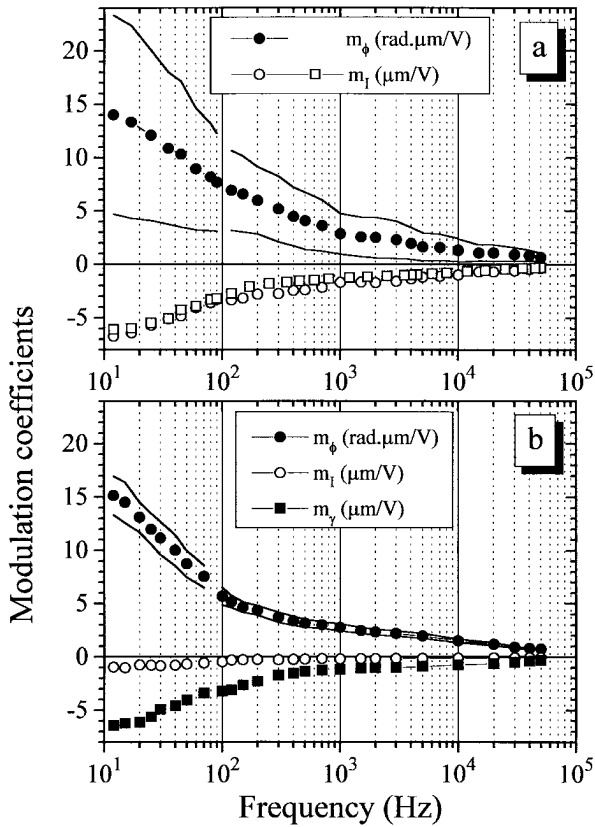

Fig. 5. Frequency dispersion of the modulation coefficients in RbHSeO$_4$. (a) Phase modulation coefficient $m_\phi$ (full circles) and intensity modulation coefficient $m_I$ (open circles) recorded in the first experiment, with a diaphragm selecting the direct beam. Open squares are direct measurements of $m_I$ performed independently (without an analyzer). (b) Phase modulation coefficient $m_\phi$ (full circles), intensity modulation coefficient $m_I$ (open circles), and contrast modulation coefficient $m_\gamma$ (full squares) recorded in the second experiment (without a diaphragm, beams $D$, $A$, and $B$ together). On both plots, the full lines correspond to the erroneous results which would be obtained for $m_\phi$ from single measurements at only one middle point (either $M_1$ or $M_1'$), $m_I$ being ignored.

During the experiments, we observed that the first double-frequency point was shifted by only a few degrees from the point $M_0$ of minimal intensity, while the second double-frequency point experienced a greater shift from the point $M_0'$ of maximal intensity (typically 60° instead of 90°). These values are fairly consistent with what can be calculated from (11), neglecting the contrast modulation. This means that mainly intensity modulation is responsible for these shifts: contrast modulation on the direct beam is actually weak. This feature is worth discussing. The loss of contrast on the direct beam ($1 - \gamma_D = 10\%$ in this experiment) can be mainly attributed to:

- Imperfect parallelism of the faces of the sample.
- Walk-off of the extraordinary beam.
- Multiple deflection processes $AB$ and $BA$ (as well as $R^2$, $A'B'$, $B'A'$, $RAB'$, $\cdots$ and their algebraic products) which lead to incoherent rays parallel to the direct rays of the $D$ beam.
- Distribution of the optical pathlengths followed by the different direct rays through the domain structure.

The first two causes are probably predominant but they are not—or are weakly—sensitive to the modulation. The third cause could give rise to a small contrast modulation, but the order of magnitude should be small, approximately $m_\gamma \approx m_I \times 2I_A I_B/I^2 \approx 0.08 m_I$ in our experiment. The fourth cause could be important if domains were scarce, that is, if the crystal were brought close to ferroelectric saturation by a dc field. Since our experiments were done in the remanent state ($E_{\rm dc} = 0$), most probably the domain structure was dense enough to induce a significant but field-insensitive loss of contrast.

*2) Second Experiment: Beams $D$, $A$, and $B$ Together:* When the three beams are collected by a lens onto the photodetector, intensity modulation is much weaker than in the first experiment, as evidenced in Fig. 4(b): the single-frequency signals recorded at the middle points of the transfer function have similar amplitudes. The slight residual intensity modulation can be attributed to the retro-deflected beams $R$, $A'$, and $B'$ [Fig. 3(b)] which were not collected during this experiment. On the other hand, the contrast is weak ($\gamma \approx 51\%$) and the double-frequency points are noticeably shifted from the extremal points. ($\beta_{2\omega} - \beta_0 \approx -12°$ and $\beta'_{2\omega} - \beta'_0 \approx -16°$.) The quasi-equality of these shift angles reflects the weakness of the intensity modulation, but their magnitude indicates a large contrast modulation. We have measured the amplitude of the modulated signal at the middle points of the transfer function, as well as the shift angles of the double-frequency points, for several frequencies in the range 10 Hz–50 kHz. The modulation coefficients $m_\phi$ and $m_I$ have been determined from (15) and (16). Then, using (17), we have deduced the contrast modulation coefficient $m_\gamma$ from the shift angles $\Delta\beta^+$ and $\Delta\beta^-$ for each frequency. The results are plotted in Fig. 5(b).

*3) Comparison of the Two Methods:* Looking at the experimental results of Fig. 5, two remarks can be made.

- For the phase modulation coefficient $m_\phi$, the dispersions obtained in both experiments are in fair agreement.
- The dispersions of the three modulation coefficients $m_\phi$, $m_\gamma$, and $m_I$ are similar and, moreover, the coefficient $m_I$ (first method) and $m_\gamma$ (second method) are nearly equal in the whole frequency range of the measurements.

These features can be readily explained by the fact that both additional modulation effects (intensity modulation in the first method, contrast modulation in the second method) are actually one single and unique phenomenon. In the second experiment, the deflected light is added to the $D$-beam, but the absolute range $I_{\max} - I_{\min}$ of the transfer function is evidently the same as in the first experiment, because the deflected light is incoherent. Therefore,

$$\gamma_D I_D = \gamma I. \qquad (18)$$

Since both the contrast $\gamma_D$ of the $D$-beam and the total intensity $I = I_D + I_A + I_B$ are nearly insensitive to the electric field, the derivative of (18) versus $E$ yields immediately

$$\frac{1}{I_D}\frac{dI_D}{dE} \approx \frac{1}{\gamma}\frac{d\gamma}{dE} \Rightarrow m_I \approx m_\gamma. \qquad (19)$$

In brief, it can be said that from the first experiment to the second one, intensity modulation has turned into contrast modulation. Correlatively, there is no change for $m_\phi$, since the deflected light is insensitive to phase modulation.

Ultimately, it is worth discussing whether the first or the second method is better to measure the EO coefficient. It could be believed that in the second experiment, since the intensity modulation is very weak, the phase modulation coefficient is determined more easily—and perhaps more accurately—than in the first experiment. This is certainly true when the intensity modulation is stronger—or nearly as strong as—the phase modulation. However, a disadvantage of the second method is that the average contrast is obviously weaker. This can increase the experimental error on the EO coefficient, even dramatically when the contrast is very weak. With some RHSe crystals, we observe such a strong deflection—especially when the sample is brought close to the coercive state—that the contrast of the total transmitted light can fall down to less than 30%, while it is usually in the range 60%–95% on the $D$-beam alone (depending on the samples and on the experimental conditions).

In both methods, the shift of the double-frequency points can be somewhat disturbing for the experimenter, who has to bear in mind that the single-frequency signal should always be measured not at 45° from a double-frequency point—as is usually done—but at the middle points of the transfer function. This can be troublesome if the measurements are performed with an oscilloscope in ac mode, especially when the stability of the transfer function is perturbed in time by some external factors, such as thermooptic effects or slow recovering of the sample in a new ferroelectric state after a step of dc field.

For all these reasons, we cannot prescribe one method or the other. For multidomain ferroelectric materials, it is recommended to use the first method when the contrast is poor (direct beam selected by a diaphragm). The second method (total light collected by a lens) may be preferred when the contrast is good. In any event, it is necessary to perform the double measurement at both linear points of the transfer function and to always check that the average intensity measured at these points is equal to $(I_{\max} + I_{\min})/2$; *when additional modulation effects are superimposed to the phase modulation, the double-frequency points can no longer be used as reliable references to find the linear points.*

## V. Consequences for Electrooptic Modulators

As a rule, the quality of the contrast is most often considered as one of the main requirements for electrooptic modulators. Consequently, in the case of scattering materials such as ferroelectric ceramics or multidomain crystals, it is, of course, recommended to select the direct beam and to hide the incoherent light by a diaphragm at the output of the sample. As a result, the output signal usually combines phase modulation and intensity modulation. If the modulator is devoted to applications in phase modulation only, the intensity modulation itself is evidently useless, but it does not affect the phenomenon. On the other hand, if the modulator is devoted to applications in intensity modulation, it could be important to choose conveniently the linear working point (either $M_1$ or $M_1'$) of the modulator in order to enhance the overall amplitude of the signal. This is illustrated in Fig. 6: the upper curve on each plot is the output signal given by a modulator working at its own half-wave voltage between polarizers in a Sénarmont's setup, with a contrast of 90% for the transfer function and with a significant intensity modulation superimposed to the phase modulation ($m_\phi/m_I = 2\pi$ rad). In Fig. 6(a), at the working point $M_1'$, where intensity modulation is working against phase modulation, the output signal is deformed and its amplitude is not optimized. It is obviously more convenient to choose $M_1$ as a working point [Fig. 6(b)] since the amplitude of the output signal is enhanced and the effective contrast is better. Accordingly, this fact must be considered when building or using electrooptic modulators made from deflecting or diffusive materials, such as multidomain crystals (RHSe) or semitransparent ceramics; for applications in intensity modulation, it is not always equivalent to change the angular position of any polarizer by 90°. In other words, it is not surprising that the "normally on" and the "normally off" setups, working under the same half-wave voltage $(0/V_\pi)$, could give more or less different signals, as soon as the fundamental component of the direct intensity modulation is not negligible.

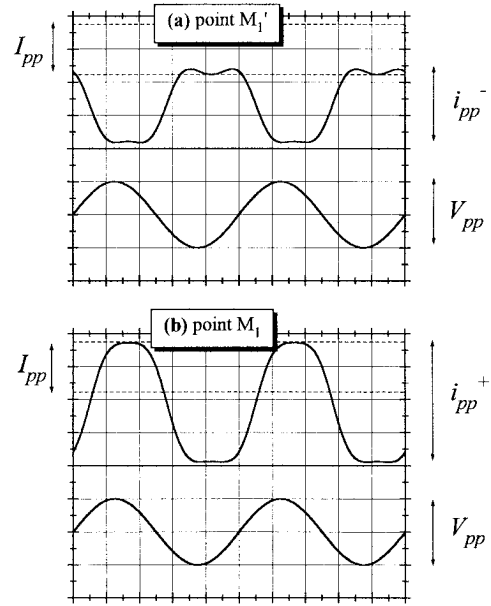

Fig. 6. Simulation of the single-frequency signals given by a modulator working at its half-wave voltage in the Sénarmont's setup, assuming a contrast of 90% and a significant intensity modulation superimposed on the phase modulation ($m_\phi/m_I = 2\pi$ rad). $I_{\rm pp}$ is the amplitude of the direct intensity modulation ($I_{\rm pp} = \gamma \cdot m_I \cdot I_M \cdot E_{\rm pp}$). On both plots, the baseline of zero intensity for the optical signal (upper signal) is the midline of the screen.

## VI. Conclusion

We have shown that additional modulation effects—intensity and/or contrast modulation—can disturb the phase modulation in electrooptic experiments. These effects can be strong in multidomain ferroelectric materials, where the intensity of the deflected or scattered light is generally sensitive to the electric field. To conveniently discriminate the different contributions, experimental and numerical procedures have been explained and compared. The consequences of the EO modulators have been examined. It should be noticed that we focused our study on the Sénarmont's setup. Nevertheless, the additional modulation effects that we discussed above can be a source of error or signal perturbation in any electrooptic setup based on electrically induced intensity variations and from which a modulated phase variation is measured, such as a two-wave setup using a Michelson or a Mach–Zender interferometer.

Finally, it is worth recalling that we have considered in the calculations the simple case of EO materials where the combined modulation effects have the same physical origin. As already mentioned, any difference between the phase retardations associated to each contribution, implying that the corresponding modulation coefficients can no longer be considered as real quantities. In such a case—for instance, when phase modulation is mainly due to ionic or electronic Pockels contributions while intensity modulation is due to domain dynamics or indirect effects—the optical signal is not only modified in amplitude but also distorted from the sinus shape. Accordingly, a more careful analysis of the signal is required to extract the different contributions in this case (using for instance a lock-in amplifier instead of an oscilloscope or a multimeter). The above calculations could nevertheless be transposed with the modulation coefficients considered as complex quantities. If this refined numerical treatment is not applied to determine the EO coefficients, we recommend using the second experimental method rather than the first one, i.e., to turn intensity modulation into contrast modulation by replacing the diaphragm with a lens at the output of the sample. This advice may seem surprising, since the contrast will be weaker, but in this way the single-frequency signal will be less distorted from the sinus shape and the determination of the EO coefficient will be more straightforward.

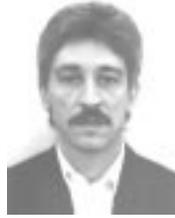

**Laurent Guilbert** was born in France in 1959. He received the Ph.D. degree in solid-state physics from the University Pierre & Marie Curie (Paris VI), France, in 1991.

He performed thesis work at Thomson Laboratories (Laboratoire Central de Recherches Thomson CSF, Orsay) from 1986 to 1991. He held temporary positions in several French universities (Le Mans and Lille) and then became Matre de Conferences at the University of Metz, Metz, France, in 1993. His research until 1992 was devoted to various topics in solid-state physics: crystal growth of gallium arsenide by the Czochralski technique and the study of grown-in defects, vibrational and magnetic properties of rare-earth fluorides, and electrical and optical properties of dislocations in III–V heterostructures. Since 1993, he has been a Researcher with Laboratoire Matriaux Opiques Proprits Spcifiques, University of Metz. He turned to experimental and theoretical studies in optical materials, with special interest in electrooptics, light deflection, and nonlinear effects in hydrogen bonded crystals exhibiting layered domain structures.

Dr. Guilbert is a member of the Société Française d'Optique.

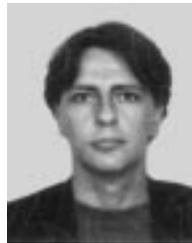

**Jean Paul Salvestrini** was born in 1966. He received the Ph.D. degree from the University of Metz, Metz, France, in 1995. His dissertation was entitled "Electro-optic modulation figures of merit of rubidium and ammonium hydrogen selenates."

Since 1995, he has been working, as Matre de Confrences, at the Laboratoire Matriaux Optiques Proprits Spcifiques, which is part of the Centre Lorrain d'Optique et d'Electronique des Solides, University of Metz. His current research activities deal with experimental and theoretical studies in electrooptics (Pockels effect, Q-switch), light deflection, and nonlinear optical effects (second-harmonic generation) in inorganic and hybrid organic–inorganic crystals. DHe holds three patents.

Dr. Salvestrini is a fellow of the Société Française d'Optique and the European Optical Society.

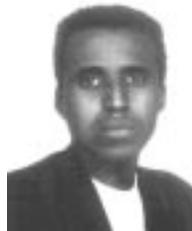

**Houssein Hassan** was born in Dirri-Daoua, Ethiopia, on July 29, 1964. He received the M.Sc. degree in engineering physics from the University of Metz, Metz, France, in 1989 and the Ph.D. degree in 1995 from the Centre Lorrain d'Optique et d'Electronique (C.L.O.E.S) of the University of Metz. His research was devoted to the study of phase transitions in PLZT ceramic system by means of Raman and dielectric spectroscopy.

In 1996, he joined the team of Dr. Fontana, Dr. Guilbert and Dr. Salvestrini, where his research has been concerned with nonlinear optics, mainly in PZT and PLZT ceramics, for such applications as switching and modulation.

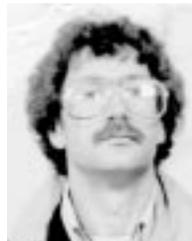

**Marc D. Fontana** was born in France in 1953. He received the M.S. and DEA degrees in physics from the University Louis Pasteur, Strasbourg. His thesis focused on the lattice dynamics and phase transitions in $KNbO_3$. He received the Ph.D. degree from the University of Metz, Metz, France, in 1985 for his work on soft mode and critical fluctuations in the successive phase transitions of $KNbO_3$ and $K(Ta,Nb)O_3$.

He was Assistant Professor at the University of Metz. Then, after a post-doctoral position at IBM Laboratories, Ruschlikon, Switzerland, his research was devoted to the studies of dielectric and lattice dynamical properties of various oxides. He became Professor at the University in 1988 and his interest turned to the measurement and the mechanisms of electrooptic properties in inorganic materials. He is the author of more than 100 papers. He is head of the Laboratoire Matériaux Optiques à Propriétés Spécifiques, Centre Lorrain d'Optique et d'Electronique des Solides, University of Metz, which he is concerned with the preparation and characterization of optical (mainly inorganic dielectric) materials for their use in nonlinear optics and electrooptics.